# Spin stripe order in a square planar trilayer nickelate

Junjie Zhang[1,2], D. M. Pajerowski[3], A. S. Botana[1], Hong Zheng[1], L. Harriger[4], J. Rodriguez-Rivera[4,5], J. P. C. Ruff[6], N. J. Schreiber[7], B. Wang[1], Yu-Sheng Chen[8], W. C. Chen[4,5], M. R. Norman[1], S. Rosenkranz[1], J. F. Mitchell[1], and D. Phelan[1,*]

[1]Materials Science Division, Argonne National Laboratory, Argonne, Illinois 60439, USA

[2]Materials Science and Technology Division, Oak Ridge National Laboratory, Oak Ridge, Tennessee 37831, USA

[3]Quantum Condensed Matter Division, Oak Ridge National Laboratory, Oak Ridge, Tennessee 37831, USA

[4]NIST Center for Neutron Research, National Institute of Standards and Technology, Gaithersburg, Maryland 20899, USA

[5]Department of Materials Sciences, University of Maryland, College Park, Maryland 20742, USA

[6]CHESS, Cornell University, Ithaca, NY 14853, USA

[7]Department of Materials Science and Engineering, Cornell University, Ithaca, NY 14853, USA

[8]ChemMatCARS, The University of Chicago, Argonne, IL 60439, USA

**Abstract:**

Trilayer nickelates, which exhibit a high degree of orbital polarization combined with an electron count ($d^{8.67}$) corresponding to overdoped cuprates, have been identified as a promising candidate platform for achieving high-$T_c$ superconductivity. One such material, La$_4$Ni$_3$O$_8$, undergoes a semiconductor-insulator transition at ~105 K, which was recently shown to arise from the formation of charge stripes. However, an outstanding issue has been the origin of an anomaly in the magnetic susceptibility at the transition and whether it signifies formation of spin stripes akin to single layer nickelates. Here we report single crystal neutron diffraction measurements (both polarized and unpolarized) that establish that the ground state is indeed magnetic. The ordering is modeled as antiferromagnetic spin stripes that are commensurate with the charge stripes, the magnetic ordering occurring in individual trilayers that are essentially uncorrelated along the crystallographic $c$-axis. Comparison of the charge and spin stripe order parameters reveals that, in contrast to single-layer nickelates such as $La_{2-x}Sr_xNiO_4$ as well as related quasi-2D oxides including manganites, cobaltates, and cuprates, these orders uniquely appear simultaneously, thus demonstrating a stronger coupling between spin and charge than in these related low-dimensional correlated oxides.

**Main text:**

There has been intense interest in stripe phases due to the interplay of charge, spin and lattice degrees of freedom as well as their relevance to high-temperature superconductivity in cuprates [1-9]. Uncovering cuprate-like superconductivity in oxides containing transition metals other than copper remains a daunting challenge [10], and in this regard $R_4Ni_3O_8$ (R=La, Pr, or Nd) compounds have emerged as potential candidates [11-13]. These layered materials possess structures that resemble the $n$=3 Ruddlesden-Popper phase ($R_{n+1}Ni_nO_{3n+1}$) [14], but they differ in that all apical oxygens are absent, resulting in trilayers of $NiO_2$ planes in which all Ni ions possess square-planar coordination of oxygen anions. The electron count ($3d^{8.67}$) coincides with the over-doped regime of cuprates [12,15]. Recent work indicates that these nickelates possess a low-spin state of Ni, large orbital polarization of the $e_g$ states with predominantly $d_{x^2-y^2}$ orbital character near the Fermi energy, and significant O $2p$-Ni $3d$ hybridization, all of which are considered to be important ingredients for superconductivity in the high-$T_c$ cuprates [12]. Thus $R_4Ni_3O_8$ compounds (particularly $Pr_4Ni_3O_8$ which is metallic in its ground state [12]) are more similar to the superconducting cuprates than previously studied nickelates with octahedral coordination, such as $La_{2-x}Sr_xNiO_4$ (LSNO) [16-18] and $LaNiO_3$-based heterostructures [19].

Unlike metallic $Pr_4Ni_3O_8$, $La_4Ni_3O_8$ undergoes a semiconductor-insulator transition upon cooling through ~105 K [11,13,20-28], and we have recently shown that the insulating state is characterized by the formation of charge stripes [13]. These stripes form in the Ni-O planes and are oriented at 45° to the Ni-O bonds, tripling the unit cell along the propagation direction. This tripling can be modeled as an ordering of charges in a 2:1 ratio (consistent with a splitting of the nominal charge concentration ($Ni^{4/3+}$) into $Ni^{1+}$ and $Ni^{2+}$ ions), and the ensuing three-fold superlattice that forms is similar to that found in single-layer LSNO ($x$=1/3) [29]. A fundamental, unanswered question remains, however, as to whether the ground state is magnetically ordered or not. Neutron powder diffraction experiments found no evidence for magnetic Bragg reflections [11], which could reflect either lack of magnetic ordering or an ordered moment that is simply too weak to detect by powder diffraction. On the other hand, $^{139}$La nuclear magnetic resonance measurements have revealed dramatic spectral changes that were attributed to low-energy antiferromagnetic correlations [20] and argued to be associated with the onset of long-range magnetic order below 105 K. Additionally, recent density functional theory (DFT) calculations were consistent with a charge and spin-stripe ordered ground-state [27]. In this Letter, we report direct and unambiguous evidence from single crystal neutron diffraction measurements (both polarized and unpolarized) that the Ni-O trilayers are indeed magnetically ordered due to the formation of antiferromagnetic spin stripes. We find that the neutron intensity can be modeled with *ab*-plane spin stripes formed within trilayer blocks, which are uncorrelated along the *c*-axis. These spin stripes are commensurate with the charge stripes, and both form simultaneously at the insulator-metal transition below ~105 K, a distinct contrast to charge and spin stripe ordering in LSNO, other quasi-2D oxides, and more generally the vast majority of transition metal oxides.

Neutron diffraction measurements were performed on a single crystal (~50 mg) that was obtained via reduction of a Ruddlesden-Popper La$_4$Ni$_3$O$_{10}$ crystal that had been grown by the floating-zone method under high-pO$_2$ [13]. The reduction process left the crystal brittle and with a large, structured mosaic (~ 7° at full-width-half-maximum). The crystal was encapsulated in Cytop CTL-809M [30], which is an amorphous fluorinated epoxy that is commonly employed to avoid the incoherent background from hydrogen-containing epoxies. Nevertheless, $|\vec{Q}|$-dependent backgrounds were observed, likely arising from quasi-elastic scattering from the epoxy, and we have therefore subtracted backgrounds measured above 105 K to reveal the intrinsic signal from the crystal. Unpolarized measurements in the (*hk*0) scattering plane were performed on the MACS cold neutron triple-axis at the NIST Center for Neutron Research (NCNR) with $\lambda$= 4.05 Å. Polarized measurements in the (*h*0*l*) plane were performed on MACS using $^3$He polarizers with the neutron spin polarization oriented out of the plane (along the [010]) direction with fixed $E_i$=$E_f$=5.0 meV. For the polarized measurements, MACS was operated in single-detector triple-axis mode. Unpolarized measurements in the (*h*0*l*) scattering plane were performed on the triple-axis HB-1A at HFIR with $\lambda$= 2.37 Å with collimations of 48'-40'-40'-120'. High-resolution single crystal x-ray diffraction experiments were performed on a 0.6 mg single crystal in a displex at Beamline A2 at the Cornell High Energy Synchrotron Source (CHESS) ($\lambda$=0.363803 Å). Throughout this Letter, we employ a pseudo-tetragonal notation (space group *F*4/*mmm*) with lattice constants of *a*=*b* ~ 5.6 Å and *c*~26.1 Å for which the $\vec{a}$ and $\vec{b}$ axes are oriented 45° to the Ni-O bonds and correspond to the propagation directions of the charge stripes [13] (see SI Fig. 1 for a diagram that relates these two cells).

We begin by considering the reciprocal space map of the (*hk*0) scattering plane obtained at base temperature (1.8 K) with 120 K data subtracted as a background, as shown in Fig. 1A. The measurements reveal the presence of weak peaks located at (2/3, 0, 0), (4/3, 0, 0), (1, 1/3, 0), and at other, symmetrically equivalent positions in the scattering plane. These peaks occur at the same points as reported via single-crystal x-ray diffraction and correspond to the positions assigned to charge stripes [13]. Since the charge peaks are sharp in *h* and *k* in x-ray diffraction [13], there is not contamination from a diffuse nuclear component in this plane. Since neutrons are not directly sensitive to modulations in charge, the results imply one of three possibilities: (i) the scattering is of nuclear origin arising from atomic displacements that follow the charge stripe modulation; (ii) the scattering is of magnetic origin, such as from spin-stripes; or (iii) both (i) and (ii). In single layer (e.g., LSNO, $x \leq 0.5$) nickelates [7], $\vec{q}_s = 1/2 \vec{q}_c$, where $\vec{q}_s$ and $\vec{q}_c$ are the spin-stripe and charge-stripe wave-vectors, respectively. $\vec{q}_c$ is measured from the $\Gamma$ point, whereas $\vec{q}_s$ is measured from the Neel antiferromagnetic wave-vector (e.g. 100, which corresponds to the $(\pi,\pi)$ point in the $I4/mmm$ setting with $a=b\sim 3.9$ Å). Thus for the present case of La$_4$Ni$_3$O$_8$, where $\vec{q}_c = (2/3,0,0)$, then $\vec{q}_s = (1/3,0,0)$ and the charge-stripe and spin-stripe reflections ($\vec{\tau}_c$ and $\vec{\tau}_s$) coincide, e.g., at $\vec{\tau}_c = (0,0,0) + (2/3,0,0)$ and $\vec{\tau}_s = (1,0,0) - (1/3,0,0)$, making scenarios (i), (ii), and (iii) all possible.

To distinguish among these three possibilities, we measured the $l$-dependence of the neutron peaks. Fig. 1B shows a scan, performed with unpolarized neutrons, along $(4/3,0,l)$. A weak, broad peak is observed that is centered at $l=0$, and stronger broad peaks are present at $l=\pm 4$. This behavior differs from that observed in the x-ray measurements, in which pseudo-triplets centered at $l=8n$ (e.g. $l=-1,0,1$ and $l=7,8,9$) because the repeat distance for charge along $\vec{c}$ is the nearest distance between Ni-O planes, which is $c/8$ [13]. The difference in the $l$-dependence arises from the magnetic contribution to the neutron cross-section, and the fact that the peaks are centered at $l=4n$ establishes a nearest-neighbor antiferromagnetic interaction along $\vec{c}$ so that the repeat distance is twice that between nearest-neighboring Ni-O planes, i.e., $c/4$. This realization, combined with the broad nature of the peaks, suggestive of short-range correlations, led us to consider magnetic correlations within individual, uncorrelated trilayers as the origin of the scattering.

To further establish the presence of magnetic ordering, we performed neutron polarization analysis. All coherent nuclear scattering is non-spin-flip (NSF). Therefore, if a peak at a particular $\vec{Q}$ is observed in the spin-flip (SF) cross-section, but not in the NSF cross-section, this is immediately indicative of a magnetic origin. Fig. 2A shows the SF cross-section of a scan through (4/3,0,4), which is a peak that we identified as magnetic above based on the fact that it appears at $l$=4. A peak is clearly visible at 1.5 K, but is not present at 120 K. Moreover, as shown in Fig. 2B, a peak is absent in the NSF cross-section at both 1.5 K and 120 K. Therefore, the peak at (4/3,0,4) has a magnetic origin. We performed additional polarized measurements on the (4/3,0,0) peak, and the measurements were consistent with contributions from both nuclear and magnetic order (See SI Section).

Having established the presence of magnetic ordering, we note that by analogy with the single layer nickelates, $\vec{q}_s=\frac{1}{2}\vec{q}_c$ implies the formation of antiferromagnetic spin stripes in the NiO$_2$ planes that are commensurate with the charge stripe order. Using the charge stripe structure determined from x-ray diffraction as a starting-point [13], as shown in Fig. 3, we have constructed models of antiferromagnetic stripes within each individual layer of a trilayer. Consistent with DFT calculations as well as x-ray absorption spectroscopy [12], the models have finite $S$=1/2 spins only on Ni$^{1+}$, whereas Ni$^{2+}$ sites have no moment. Following expectations for 180° superexchange on a square planar lattice, we have taken the nearest-neighbor Ni$^{1+}$-Ni$^{1+}$ interactions to be antiferromagnetic. Under these constraints, models for uncorrelated trilayers possess three degrees of freedom: (1) direction of the spin axis, (2) magnetic coupling between individual layers ("intra-trilayer" coupling), and (3) magnetic coupling across charge stripes - the latter two of which are depicted in Fig. 3. As detailed in the SI, models that fully explore these degrees of freedom have been investigated. Based on comparisons of the calculated ($hk$0) maps and (4/3,0,$l$) cuts to the experimental observations, all models that did not possess antiferromagnetic coupling across the charge stripes, a spin axis with a significant component perpendicular to the charge stripes, and an antiferromagnetic intra-trilayer interaction could be eliminated. This intra-trilayer antiferromagnetic interaction can be understood by direct exchange between two $d^9$ ions each possessing a hole in its $d_{x^2-y^2}$ orbital [31,32] as has been argued for $YBa_2Cu_3O_{6+x}$ [33]. The model that best matches the data is shown in Fig. 3, and the resulting ($hk$0) map and (4/3,0,$l$) cut are shown in Fig. 1C and Fig. 1D, respectively. Although quantitative agreement between the calculated and measured intensities is impossible because of the unknown nuclear contribution to the cross-section, the model qualitatively reproduces the observed pattern in the ($hk$0) plane, the width of the peaks in $l$, the observed maxima at $l=\pm 4$, and the weaker peak at $l=0$. Comparisons between the calculations for $\vec{S}//\vec{b}$ and $\vec{S}//\vec{c}$ favor $\vec{S}//\vec{c}$ because of the relative distribution of intensities in the ($hk$0) plane; specifically, as shown in the SI, the case of $\vec{S}//\vec{b}$ yields intensities that are too weak at (1,1/3,0) and (1,-1/3,0).

Models of correlated trilayers have also been considered. We found that in these correlated models, the individual building units of individual trilayers must possess the same characteristics established for the uncorrelated trilayers discussed above: antiferromagnetic intra-trilayer interactions, antiferromagnetic coupling across charge stripes, and a spin axis perpendicular to the stripe direction. Moreover, significant broadening of the observed lineshapes along $l$ must be imposed to fit the data, yielding a correlation length along $\vec{c}$ that corresponds to roughly the height of a single trilayer. Thus, although weak coupling between nearest neighbor trilayers must be present, the data can be adequately modeled using a simplified uncorrelated trilayer model (intensities for correlated trilayer models are described in detail in the SI). We note that weak coupling along $\vec{c}$ is common in related materials such as LSNO [29]. In the case of $La_4Ni_3O_8$, this can be rationalized by the large distance between successive trilayers (~ 6.5 Å), the lack of a significant super-exchange pathway connecting the trilayers, the lateral shift in the Ni positions from one trilayer to the next (as shown in SI Fig. 3) which leads to geometric frustration, and the short correlation length of charge-stripes along $\vec{c}$ [13]. Moreover, layered structures such as $La_4Ni_3O_8$ often possess stacking faults with different numbers of layers. Such intergrowth structures may also reduce the correlation between trilayer blocks.

Our finding that the ordered moment lies along the *c*-axis contrasts with the majority of quasi-2D cuprates, nickelates, manganites, and cobaltites for which the moment lies parallel to the *ab*-plane. To understand this experimentally determined magnetic easy axis and to obtain the magnetocrystalline anisotropy energy, relativistic electronic structure calculations were performed (see SI for full details). The results show that $\vec{c}$ is the spin axis, in agreement with the diffraction, with an anisotropy energy $\Delta E = E[010] - E[001] = 50$ μeV/Ni. This value is consistent with the scale of anisotropy energy in cuprates [33]. The in-plane energy difference is smaller, $E[100] - E[010] = 2$ μeV/Ni.

We have measured the temperature-dependent order parameter of the charge stripes by measuring the (13/3,3,0) superlattice reflection in x-ray measurements, as shown in Fig. 4A. We have also determined the magnetic order parameter (Fig. 4B) by measuring the temperature dependence of (4/3,0,4), which is ideal since the magnetic cross-section peaks at $l=4$, whereas the nuclear cross-section does not. The magnetic order parameter was fit to a power law, $I \propto (1-T/T_N)^{2\beta}$, for $T<T_N$ (where $T_N$ is the Neel temperature) with fixed $\beta$ of 0.125, which corresponds to a 2D Ising system [34]. Inasmuch as the fit agrees reasonably well with the temperature dependence of the intensity, this suggests the order parameter is consistent with the quasi-2D Ising model described above. Direct determination of the exponent from free fitting would require a higher density of temperature points with significantly improved counting statistics compared to what could be measured. Within the temperature resolution of the measurements, both the charge and spin order parameters become finite at the same temperature, which is consistent with the anomaly in the magnetic susceptibility (Fig. 4C). In the single-layer nickelates, the lower spin-stripe transition temperature is also manifest by a distinct change in slope in the magnetic susceptibility [17]. Consistent with the spin and charge stripes possessing the same transition temperature in $La_4Ni_3O_8$, the slope of the susceptibility, shown in Fig. 4D, evidences no such feature below 105 K.

If one of spin or charge order were dominant, then its temperature dependence should be stronger than that of the secondary order parameter, which would go as the square of the primary order parameter. However, it is clear from Fig. 4 that both the charge and spin order parameters reach their maximum value at approximately the same temperature, suggesting a scenario of strongly coupled spin and charge stripes in which neither order parameter is secondary with respect to the other. This behavior of $La_4Ni_3O_8$ contrasts with that of the single-layer nickelates, as well as that of many other transition metal oxides, for which the charge stripes order at a higher temperature than the spin stripes with the spin order having a temperature dependence consistent with its secondary nature [7]. Specific examples of other transition metal oxides that do exhibit simultaneous charge and spin stripe transitions include the 3-dimensional perovskite $Nd_{1/3}Sr_{2/3}FeO_3$ [35], but which clearly has a dominant charge order parameter instead, as well as $Nd_{1/2}Sr_{1/2}MnO_3$ [36]. Thus, the simultaneous transitions in quasi-2D materials and non-dominant order parameters appear unique to $La_4Ni_3O_8$.

A comparison of the correlation lengths in spin and charge channels may offer additional insights. Within the *ab*-plane, both the charge and magnetic correlation lengths appear as long-range ordered, with the caveat that the quality of currently available crystals makes it impossible to differentiate correlation lengths larger than several nanometers from long-range order. Energy stability arguments based on the anisotropy energies calculated in the electronic structure calculations indicate that the in-plane correlation length must be more than 2 nm (though again it could be much larger). Along $\vec{c}$, the correlation length for both charge and spin is approximately the size of an individual trilayer. We do note, however, that correlation between nearest neighbor trilayers is required to reproduce the observed superlattice pattern in x-rays [13] whereas such coupling is not necessary to reproduce the magnetic intensity observed here, though such coupling clearly cannot be ruled out.

In summary, neutron scattering measurements unambiguously establish an ordered magnetic ground state for $La_4Ni_3O_8$. The magnetic structure is an antiferromagnetic spin stripe state in which the spins are commensurate with the charge stripes. The weak (3-4 orders of magnitude smaller than nuclear - see Fig. 2) and highly diffuse (along $\vec{c}*$) magnetic reflections explain why no magnetic Bragg peaks have been observed previously in powder diffraction. Our result resolves a longstanding issue regarding the anomalous drop in magnetic susceptibility that occurs at ~ 105 K and corroborates $^{139}$La nuclear magnetic resonance measurements that have been interpreted to reflect long-range antiferromagnetic order. Indeed, the presence of multi-domain stripes with weak or little inter-trilayer coupling may explain the broad distribution of hyperfine fields that were reported [20]. Our observations provide a unified spin stripe/charge stripe coupling picture in layered nickelates and complete the parallel between single-layer nickelates and trilayer nickelates. The square-planar nature of the trilayer material, combined with its *d*-electron count and large orbital polarization, make it appealing as a possible starting point for finding cuprate-like superconductivity. The present observation of antiferromagnetic spin stripes represents yet another intriguing parallel to the cuprates, and suppression of the charge and spin stripe states via chemical substitution or electrolytic gating may represent viable means for achieving superconductivity. Clearly, the most important question in the trilayer system is whether upon charge carrier doping the physics will parallel the persistent stripe physics of single-layer nickelates or if superconductivity and related phenomenology will emerge as in the cuprates.

**Acknowledgements:**

Work in the Materials Science Division at Argonne National Laboratory (crystal growth, characterization, neutron and x-ray scattering experiments, data analysis, and theoretical calculations) was supported by the US Department of Energy, Office of Science, Basic Energy Sciences, Materials Science and Engineering Division. This research has been supported in part by ORNL Postdoctoral Development Fund by UT-Battelle, LLC under Contract No. DE-AC05-00OR22725 with the U.S. Department of Energy. This research used resources at the High Flux Isotope Reactor, a DOE Office of Science User Facility operated by the Oak Ridge National Laboratory. This work is based upon research conducted at the Cornell High Energy Synchrotron Source (CHESS) which is supported by the National Science Foundation under award DMR-1332208. Access to MACS was provided by the Center for High Resolution Neutron Scattering, a partnership between the National Institute of Standards and Technology and the National Science Foundation under Agreement No. DMR-1508249. The authors thank Wangchun Chen and Shannon Watson for their assistance with the polarized $^3$He cells on MACS.

**Figure Captions:**

**Fig. 1**: (A) Neutron scattering intensity in the ($hk0$) plane measured for $La_4Ni_3O_8$ at T=1.8 K with 120 K data subtracted. (B) Scan along (4/3, 0, $l$) at T=2.0 K with 120 K data subtracted. Uncertainties represent one standard deviation, derived from the square roots of the numbers of counts. The simulated intensities in the ($hk0$) plane as well as along (4/3, 0, $l$) for the uncorrelated trilayer model described in the text and shown in Fig. 3 are shown in (C) and (D) respectively.

**Fig. 2**: Spin polarized neutron diffraction from $La_4Ni_3O_8$ as measured on MACS in the ($h0l$) plane with neutron spin polarization perpendicular to the plane. (a) Scans along ($h$,0,0) through (4/3,0,4) in the spin-flip channel at 1.5 K (blue) and 120 K (red). (b) The same scans in the non-spin-flip channel. Uncertainties represent one standard deviation, derived from the square roots of the number of counts.

**Fig. 3**: (A) Uncorrelated trilayer model for the spin ordering, displayed layer by layer. $Ni^{1+}$ sites are shown in blue, whereas non-magnetic $Ni^{2+}$ sites are shown in red. Moments pointing into the plot are shown by ⊗, while moments pointing out of the plot are shown as ⊙. The intra-trilayer coupling as well as the coupling across stripes is labeled. (B) 3-dimensional perspective of the structure shown in (A).

**Fig. 4**: Temperature-dependence of the charge superlattice peak (13/3,3,0) measured with x-rays (A), spin superlattice peak (4/3,0,4) measured with neutrons (B), magnetic susceptibility (C), and first derivative of the magnetic susceptibility (D). Uncertainties in (B) represent one standard deviation, derived from the square root of the number of counts. The fit in (B) corresponds to a power law as described in the text.

Fig. 1

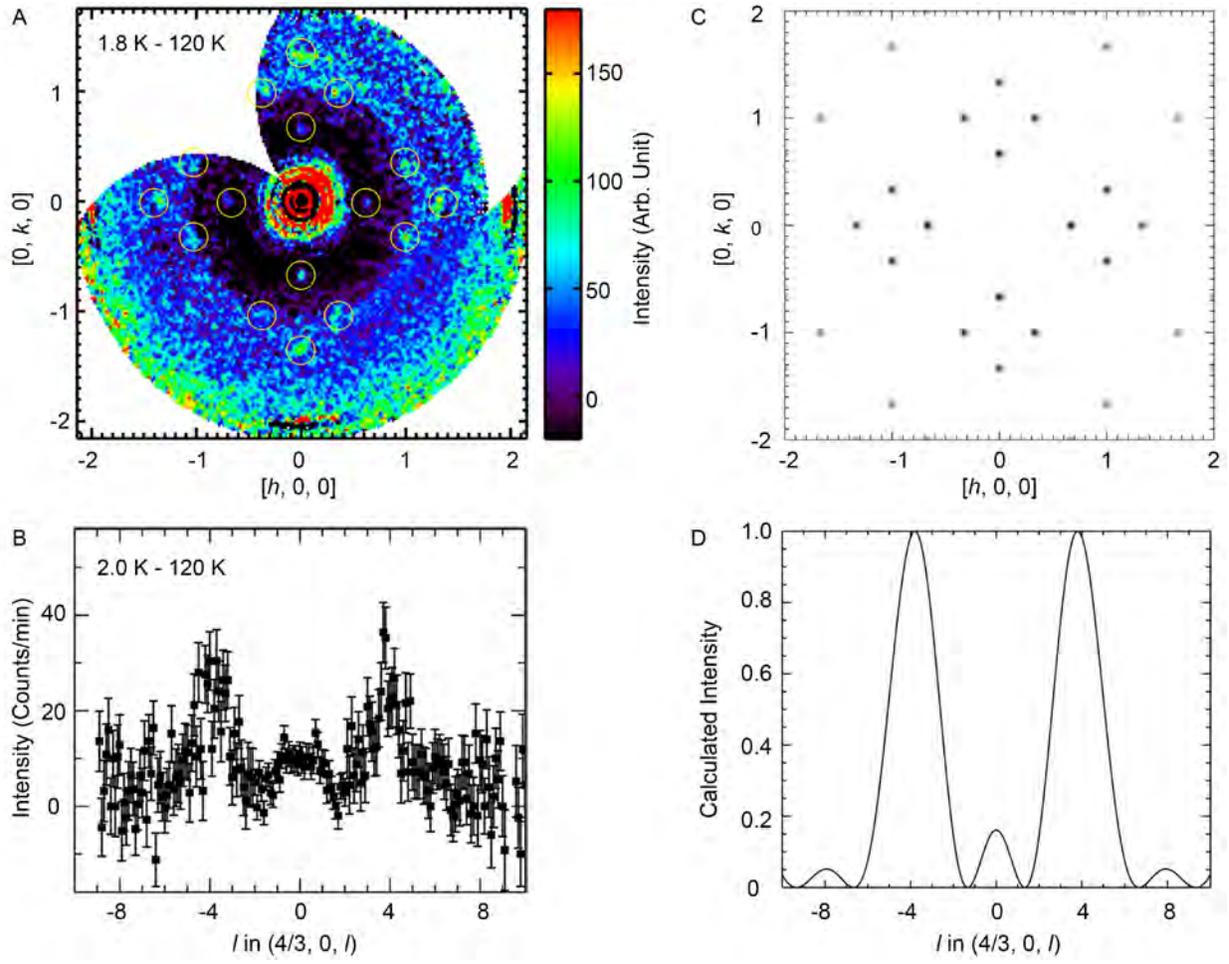

**Fig. 2**

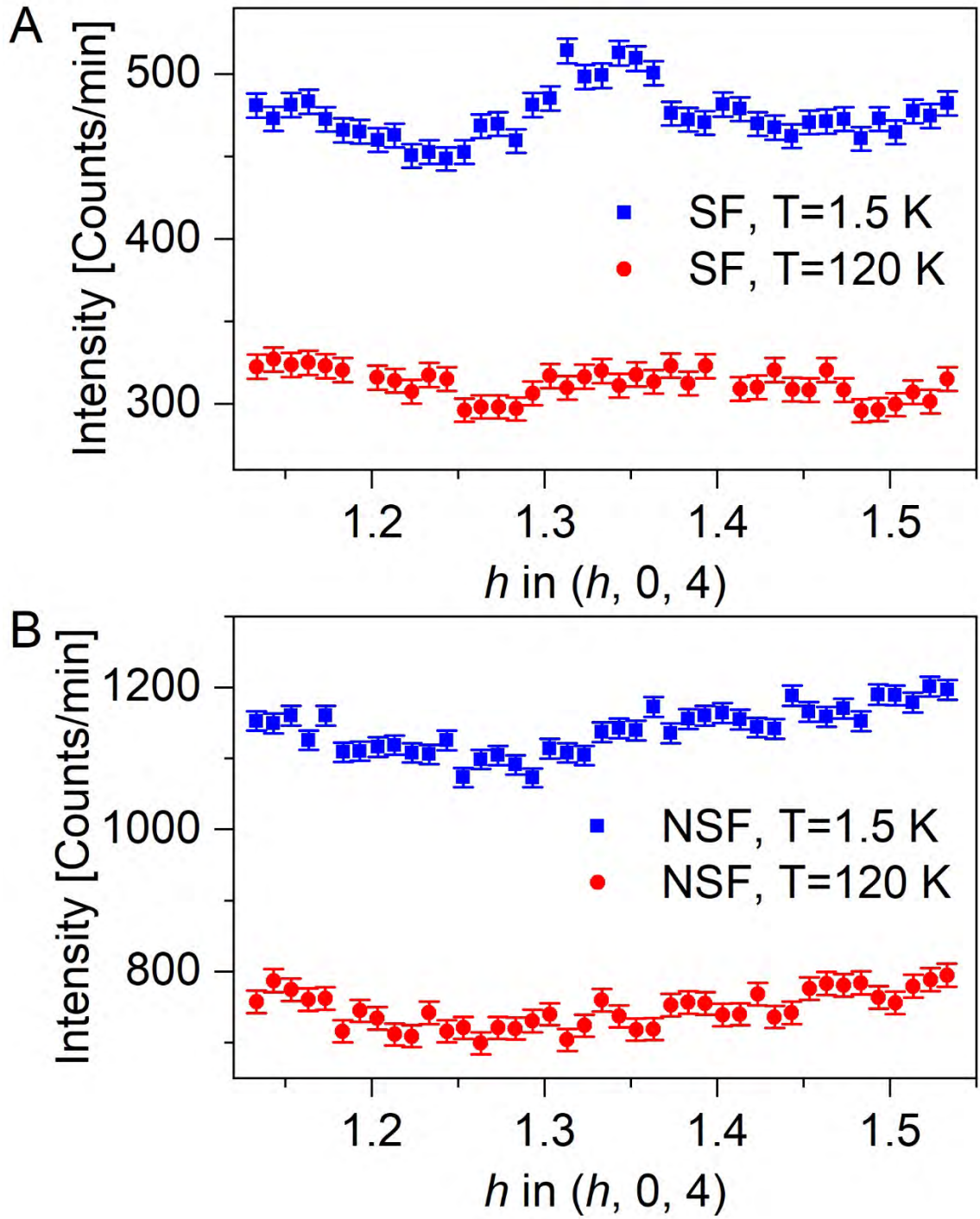

Fig. 3

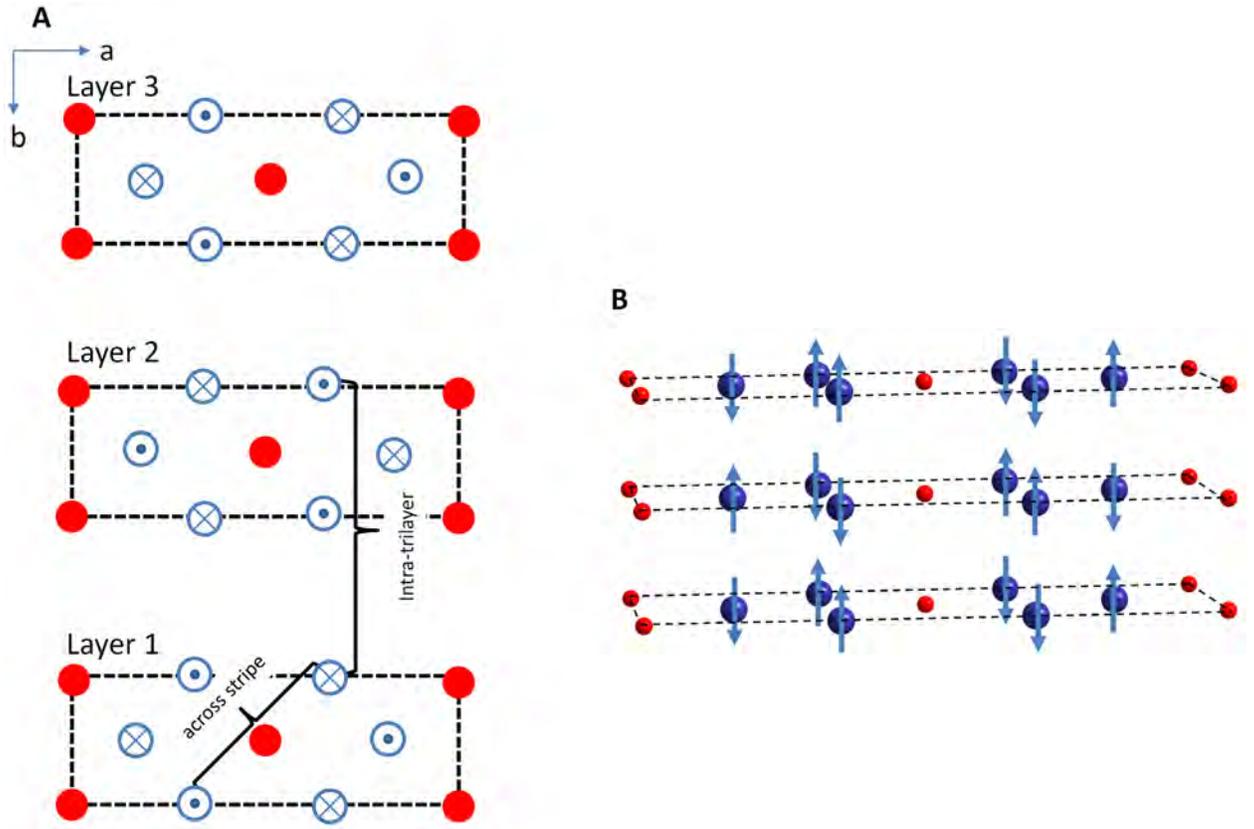

Fig. 4

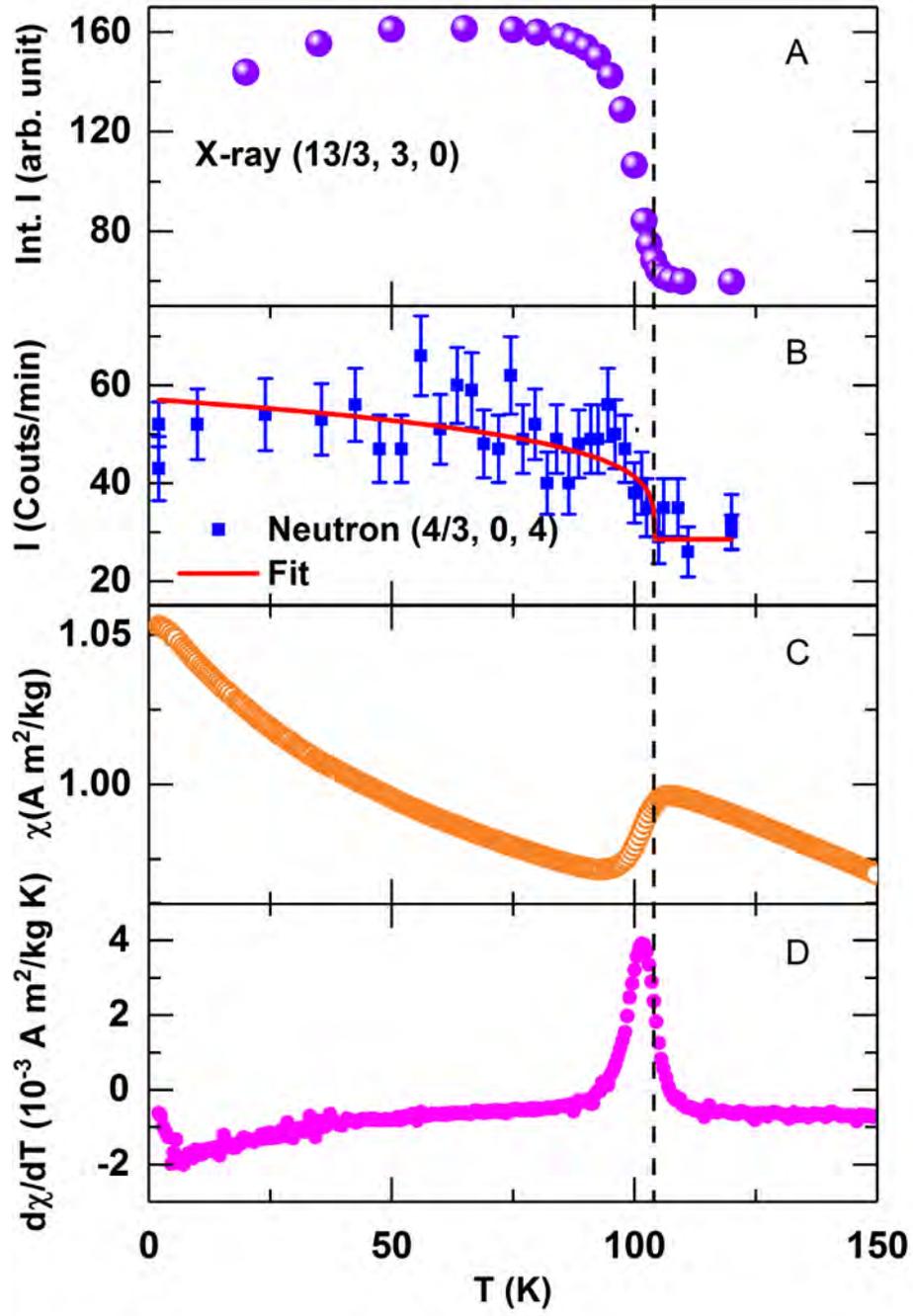

# Supplementary Information for "Spin stripe order in a square planar trilayer nickelate"

## I. Unit Cell

Throughout the manuscript, we base the description of the magnetic structure and neutron intensities upon the ~ 5.6 Å × 5.6 Å × 26.1 Å cell (space group $F4/mmm$). Here, the $\vec{a}$ and $\vec{b}$ axes are rotated by 45 degrees with respect to the Ni-O nearest-neighbor bonds. Thus $a$ and $b$ are each increased by a factor of $\sqrt{2}$ with respect to the conventional tetragonal description (space group. $I4/mmm$) that has the ~ 3.9 Å × 3.9 Å × 26.1 Å unit cell as displayed in Fig S1. In this description, the propagation vector for the stripes is aligned with the unit cell axis, which is consistent with previous work on single-layer nickelates.

## II. Additional Polarized Neutron Scattering Measurements

Polarized measurements in the ($hk$0) plane were performed on the thermal triple-axis HB-1 at the HFIR Facility at Oak Ridge National Laboratory with $\lambda$= 2.46 Å and horizontal collimations of 48'-80'-60'-240'. A Heusler monochromator and Heusler analyzer were used for neutron polarization analysis. The polarized measurements at 50 K and 300 K were performed in two separate experiments. The conditions in each case were identical, and the intensities of the nuclear Bragg peak, 220, were consistent between these two measurements.

In particular, using an in-plane vector magnet, we have oriented the neutron polarization, $\vec{P}$, parallel to the wave-vector, $\vec{Q}$. In this condition, all coherent nuclear scattering is ideally non-spin flip (NSF), while all magnetic scattering is spin flip (SF), though inefficiencies in the polarization always lead to a weak cross-contamination of the NSF and SF cross-sections. As shown in Figs. S2(a) and S2(b), at both 300 K and 50 K, respectively, rocking scans through the 220 Bragg reflection show little temperature dependence because it is completely nuclear. Thus, it is dominated by the NSF cross-section, which is ~13 times stronger than the SF cross-section. This factor (flipping ratio) defines and limits the sensitivity of our experiment.

Figs. S2(c) and S2(d) show similar rocking scans of the (4/3,0,0) reflection at both 300 K and 50 K. First, it is apparent that there is no peak in either the SF or the NSF cross-section at 300 K, showing that there is neither a charge nor spin modulation present. However, at 50 K, peaks are observed in both the SF and the NSF channels. Indeed, although the NSF intensity is a bit stronger than the SF intensity, they are comparable, and their ratio (~1.3) is an order of magnitude smaller than the flipping ratio. Thus, both nuclear and magnetic components contribute to the (4/3,0,0) cross-sections, implying that the underlying real space modulation arises from both spin-spin correlations and correlated atomic displacements at (4/3,0,0), which differs from the reflection at (4/3,0,4) which is dominated by the spin-spin component. For the case of LSNO, correlated displacements also give rise to nuclear superlattice reflections, though they do not overlap with the magnetic reflections except for the specific case of $La_{5/3}Sr_{1/3}NiO_4$ [1].

It is instructive to consider the widths of the SF and NSF rocking curves at (4/3,0,0). The widths cannot be narrower than the fundamental structural Bragg peaks which have widths that define the limit of long-range in-plane order. The broadening of the superlattice peaks beyond those of the fundamental peaks would be indicative of finite correlation lengths. We therefore compared the normalized cross-sections of the NSF and SF scans to those of the 220 structural Bragg peaks as shown in Fig. S3(a) and S3(b). While the NSF signal appears a bit narrower than the 220 rocking curve, which is physically impossible, we attribute this to the inherently low statistics of the measurement. Based on the statistics, we are unable to conclude that there is (or is not) any conclusive difference between the in-plane correlation lengths of the long-range and short-range order.

## II. Method of calculation of the magnetic intensity for uncorrelated trilayers

The magnetic neutron scattering intensity resulting from a single trilayer was determined from a 5×15×1 supercell. The structure factor was calculated from:

$$(1) \quad F_M(\vec{Q}) = \sum_n [\hat{Q}(\hat{Q} \cdot \hat{S}_n) - \hat{S}_n] |\vec{S}_n| f_n(\vec{Q}) e^{i\vec{Q} \cdot \vec{x}_n}.$$

Here, the sum is over all $n$ magnetic sites in the supercell, each site having spin $\vec{S}_n$ and position $\vec{x}_n$. $\hat{S}_n$ is a unit vector in the direction of $\vec{S}_n$. The magnetic form factor is given by $f_n(\vec{Q})$. Because of the large orbital polarization present in $La_4Ni_3O_8$, we have employed an anisotropic magnetic form factor appropriate for spins arising from an unpaired electron in the $d_{x^2-y^2}$ shell consistent with [2]. Specifically, this yields an anisotropy between the $\vec{a}^*$-$\vec{b}^*$ plane and the $\vec{c}^*$ axis, with the magnetic intensity decreasing faster with increasing $|\vec{Q}|$ in the $\vec{a}^*$-$\vec{b}^*$ plane. The simulated intensity is then given by:

$$(2) \quad I(\vec{Q}) \propto |F_M(\vec{Q})|^2.$$

$I(\vec{Q})$ was then calculated for a mesh of values in both the $\vec{a}^*$-$\vec{b}^*$ plane [i.e. ($hk$0)] as well as along (4/3,0,$l$). The simulation for ($hk$0) took into account two twins, one possessing a stripe oriented along $\vec{a}$, and the other along $\vec{b}$, assuming equal populations.

## III. Uncorrelated trilayer models for the spin ordering

We use the charge stripe model determined from single-crystal x-ray diffraction experiments as the starting point for considering the spin ordering, placing non-zero moments only on the $Ni^{1+}$ sites.  We implicitly assume that there is a unique spin axis in all of the models that are described below.  We also assume that all $Ni^{1+}$ sites possess the same magnitude of magnetic moment.  The basic building block of the ordering is an individual $NiO_2$ layer.  The nearest neighbor $Ni^{1+}$-$Ni^{1+}$ coupling within each plane of the trilayer is taken to be antiferromagnetic, mediated through oxygen, which is consistent with the spin ordering in single layer nickelates and cuprates with 180° superexchange.

Within the confines of the physically intuitive constraints described above, there are three degrees of freedom in spin ordering models of uncorrelated trilayers.  First, there is a degree of freedom of the spin axis.  Second, because of the zero moment of $Ni^{2+}$ sites, there is a degree of freedom reflecting how $Ni^{1+}$ sites are coupled when separated by a $Ni^{2+}$ site.  Third, there is a degree of freedom in the relative phase between the three layers.  Twelve models were considered that explore combinations of these three degrees of freedom, and the description of each model as well as the results from calculations using the method of calculation described above are described in Table S1.  Note that in the description of the models, the domain with charge stripes propagating along $\vec{a}$ is described. By comparison to observed intensities, all but two models could be eliminated due to inconsistencies listed in Table S1.

**Table S1:** Description of the different uncorrelated trilayer models that were tested.

| Model | Spin axis | Coupling across stripe | Intra-trilayer coupling | Result |
|---|---|---|---|---|
| 1 | $\vec{b}$ | AFM | AFM | See Supp. Fig. 4 |
| 2 | $\vec{b}$ | FM | AFM | Eliminated by ($hk0$) |
| 3 | $\vec{b}$ | AFM | FM | Eliminated by $l$-dependence |

| 4 | $\vec{b}$ | FM | FM | Eliminated by *l*-dependence |
| 5 | $\vec{c}$ | AFM | AFM | See Supp. Fig. 4 |
| 6 | $\vec{c}$ | FM | AFM | Eliminated by (*hk*0) |
| 7 | $\vec{c}$ | AFM | FM | Eliminated by *l*-dependence |
| 8 | $\vec{c}$ | FM | FM | Eliminated by *l*-dependence |
| 9 | $\vec{a}$ | AFM | AFM | Eliminated by (*hk*0) |
| 10 | $\vec{a}$ | FM | AFM | Eliminated by (*hk*0) |
| 11 | $\vec{a}$ | AFM | FM | Eliminated by (*hk*0) |
| 12 | $\vec{a}$ | FM | FM | Eliminated by (*hk*0) |

From the twelve models considered, it became clear that the coupling across the stripe is antiferromagnetic, that there is an antiferromagnetic coupling between individual layers, and that the spin axis (or at least a significant component thereof) is perpendicular to the charge stripe direction. This model is depicted in the main text Fig. 3. The remaining question is whether the antiferromagnetic axis is along $\vec{b}$ or $\vec{c}$. Fig. S4 compares both the (*hk*0) plane in panels (a) and (b), and the (4/3,0,*l*) cut in panels (c) and (d) for the spin axis pointing along $\vec{b}$ and $\vec{c}$, respectively. The major difference in these models is the relative intensities of peaks in the (*hk*0) plane. Specifically, with $\vec{b}$ as the antiferromagnetic axis, the intensities of (1,1/3,0) and (1,-1/3,0) are weaker than (2/3,0,0) and (4/3,0,0). On the other hand, with $\vec{c}$ as the antiferromagnetic axis, all four of these intensities are comparable. The latter simulations are more consistent with our observations, particularly of relatively strong intensities at (1,1/3,0) and (1,-1/3,0), which favors the model with spins along $\vec{c}$. We temper this conclusion with the knowledge that we have not strictly measured the relative magnetic structure factors due to the nuclear contribution in the experiment.

**IV. Models for the spin ordering that include trilayer coupling**

Intensities were also calculated for three-dimensional models that include coupling between trilayers. The structure factors of Bragg reflections ($\vec{\tau}_{hk}$) for the unit cell of a particular model were calculated using Equation (1) for $\vec{Q}=\vec{\tau}_{hkl}$. The intensities of a particular reflection were calculated from Equation (2), and the Bragg intensities were broadened by Gaussian functions to simulate the observed intensity in the ($hk$0) plane. For cuts along (4/3,0,$l$), where extremely broad lineshapes were observed, Lorentzian functions were used to give the peak a linewidth, the inverse of the half-width-at-half maximum corresponding to the correlation length along $\vec{c}$. It is important to note that such broadening yields a correlation length corresponding to approximately the height of an individual trilayer, and when comparing simulations below, the width has been set to exactly that corresponding to a trilayer.

In the correlated trilayer models we have considered, we implicitly assumed that the ordering motif within each trilayer is identical for a given model. We found that for correlated trilayer models, the ordering within the individual trilayers follows the same rules as for the uncorrelated models. In other words, trilayer orderings eliminated in SI Table I for the uncorrelated models can also be eliminated in the correlated trilayer models. Thus, within the trilayers, the ordering must follow Model #1 or Model #5 from SI Table I. We consider that every other trilayer (alternating trilayers) possess the same phase since there is no evidence of a magnetic unit cell that is larger than the nuclear unit cell along $\vec{c}$. Thus, constructing a model requires determination of the phases between nearest-neighbor trilayers, and there are only two possibilities, both of which are illustrated in Fig. S5. In total, then, there are four correlated models that need to be considered: Model#1 Ver. A, Model#1 Ver. B, Model#5 Ver. A, and Model#5 Ver. B. Simulated intensities in the ($hk0$) plane and along (4/3,0,$l$) are shown for each of these four models in Fig. S6.

Consistent with the uncorrelated trilayer calculation, both correlated models predict intensities at (1,1/3,0) and (1,-1/3,0) that are significantly weaker than at (2/3,0,0) and (4/3,0,0) as opposed to the experimental observation when $\vec{b}$ is the antiferromagnetic axis (i.e. Model#1 Ver. A and Ver. B), though the predicted intensity is a bit stronger for Ver. A. Moreover, for both models tested with inter-trilayer coupling of type Ver. A, the multiplet around $l$=4 is predicted to possess a characteristic splitting in its line-shape, which is not observed in the experiment. The remaining model (Model #5, Ver. B) also predicts relatively weak intensities at (1,1/3,0) and (1, -1/3, 0). Thus, unless there is a strong nuclear contamination at (1,1/3,0) which increases its strength relative to that of (2/3,0,0) and (4/3,0,0), our calculations tend to favor the uncorrelated trilayer model as opposed to any of these correlated models.

## V. Electronic structure calculations

The magnetocrystalline anisotropy energies were calculated by performing relativistic electronic structure calculations. The all-electron, full potential code WIEN2K [3] was employed based on the augmented plane wave plus local orbital (APW+lo) basis set [4]. A $3a \times a \times c$ supercell was used with the charge/spin pattern shown in Fig. 3 of the main text. The chosen exchange correlation functional is the Perdew-Burke-Ernzerhof version of the generalized gradient approximation [5] that correctly reproduces the charge stripe, insulating state [6]. Spin orbit coupling was introduced in a second variational procedure with the spin axis fixed along the [100], [010], and [001] axes. The *ab initio* determination of the magnetocrystalline anisotropy requires a dense *k*-point mesh for the Brillouin zone (BZ) integration since the energy differences being targeted are small (μeV- meV per atom) [7]. Up to 1000 k-points in the full BZ were used to reach convergence. For the $Ni^{1+}$ ions, our relativistic calculations yield magnetic moments of ± 0.6 μB and an orbital moment of 0.05 μB. For $Ni^{2+}$, the moment is zero. The magnetic and orbital moment of the oxygen atoms is zero.

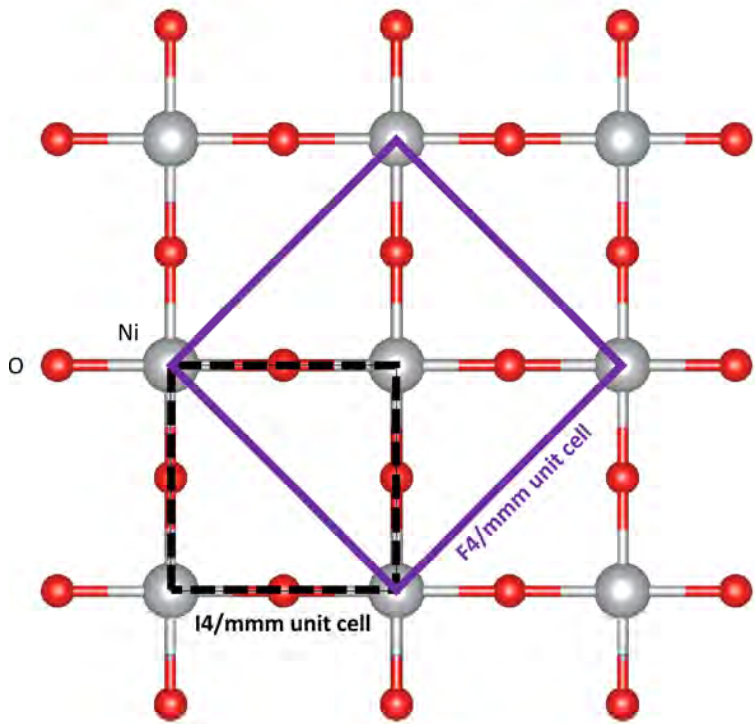

Fig. S1: Comparison of the *I*4/*mmm* and *F*4/*mmm* $\vec{a}$-$\vec{b}$ axes in a single NiO$_2$ layer.

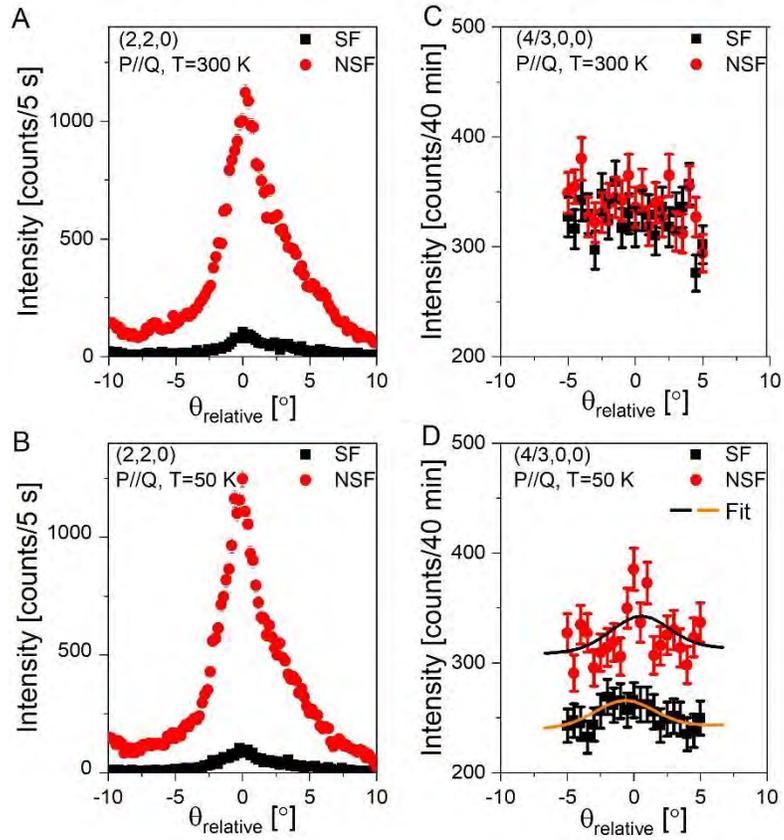

Fig. S2: Spin polarized neutron diffraction from $La_4Ni_3O_8$. Rocking scans through the 220 Bragg reflection at 300 K (A) and 50 K (B), and rocking scans of the (4/3,0,0) reflection at 300 K (C) and 50 K (D). Uncertainties represent one standard deviation, derived from the square roots of the number of counts. The non-spin-flip data in (D) has been constrained to have the same width as that obtained from the fit of the SF data.

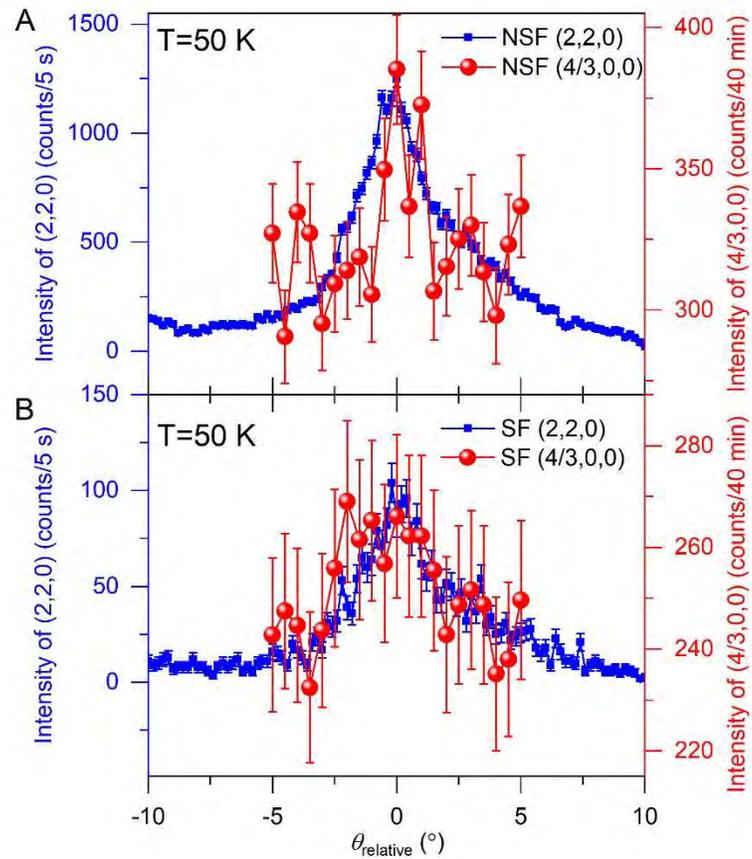

Fig. S3: (A) Non-spin-flip (NSF) rocking scan through (4/3,0,0) overplotted with the rocking scan of 220. (B) Spin-flip (SF) rocking scan through (4/3,0,0) overplotted with the rocking scan of 220.

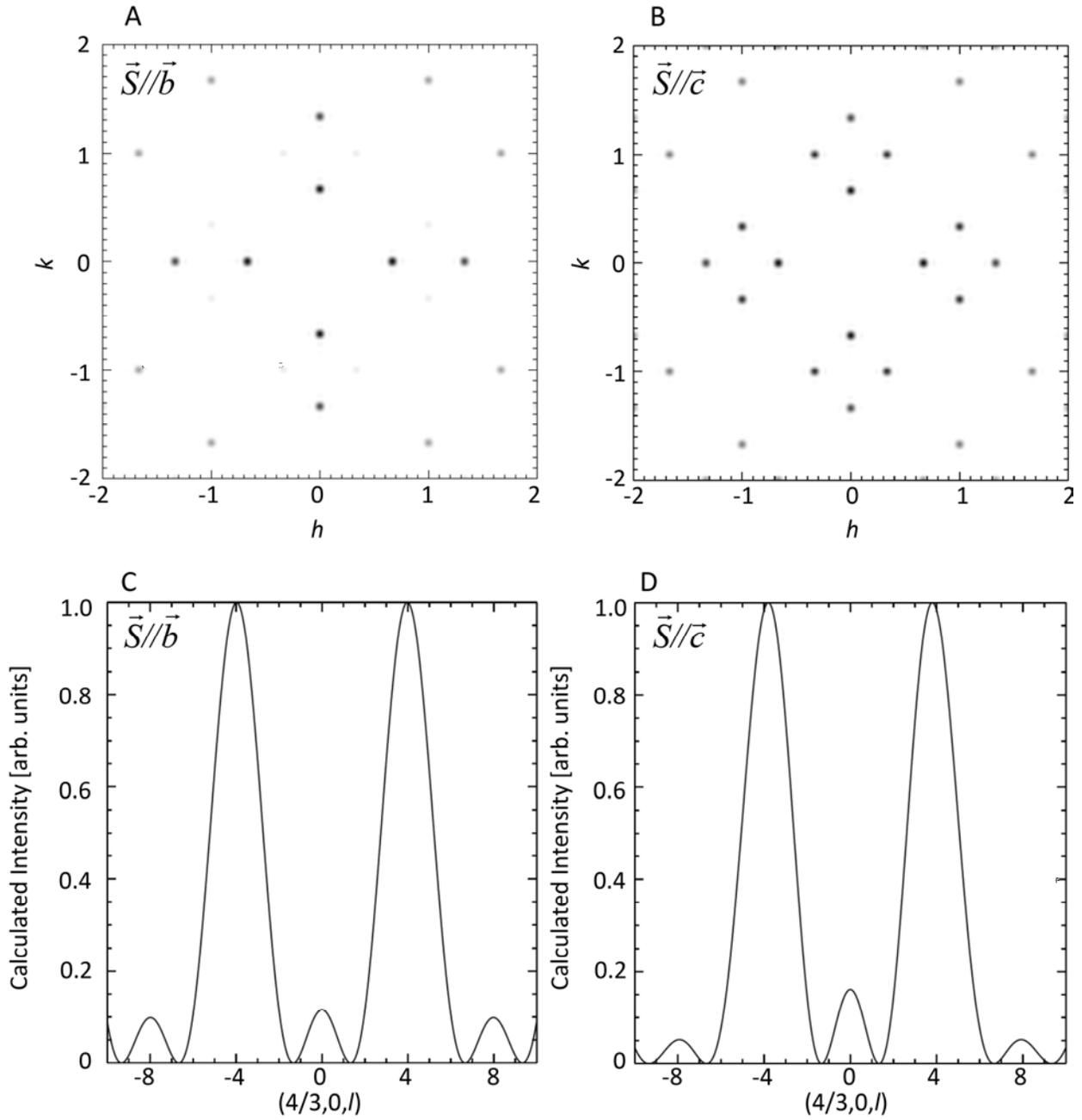

Fig. S4: Calculated intensities for uncorrelated trilayer models. The ($hk0$) plane is shown in panels A and B for $\vec{S}//\vec{b}$ and $\vec{S}//\vec{c}$, respectively. The intensity along (4/3,0,$l$) is shown in panels C and D, respectively.

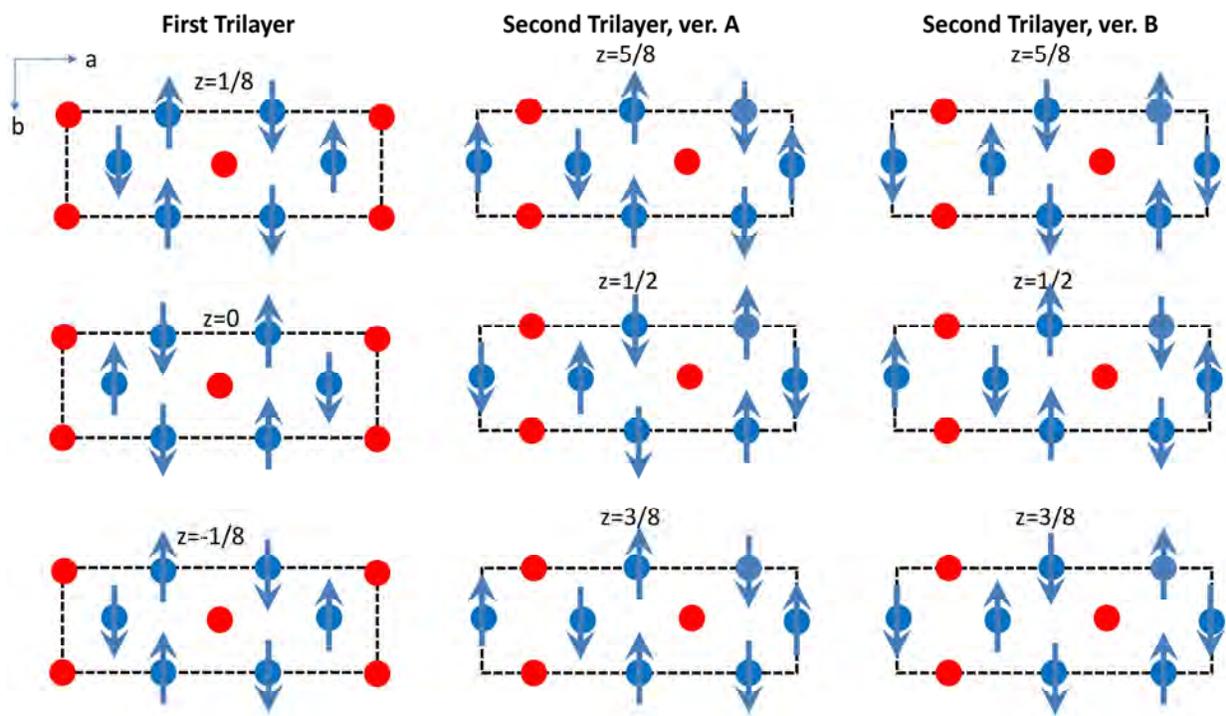

Fig. S5: Two possible models of the spin ordering. The first column shows the ordering pattern within the first trilayer. The second and third columns show the two possible ordering patterns for the adjacent trilayer.

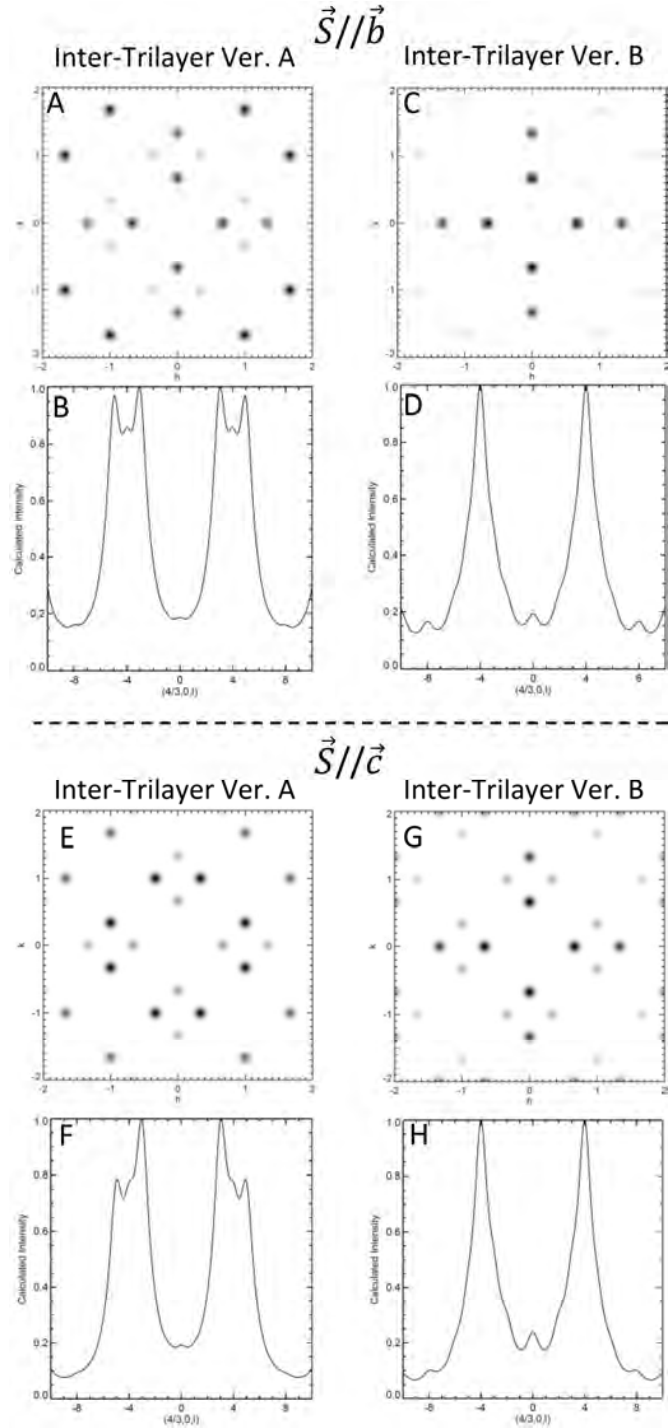

Fig. S6: Intensities calculated for the models that include coupling between trilayers. The models tested are depicted in Fig. S3 (Ver. A and Ver. B). Panels A&B and C&D show intensities in the ($hk0$) plane and along ($4/3,0,l$) with $\vec{S}//\vec{b}$ for Ver. A and Ver. B, respectively. Panels E&F and G&H show intensities with $\vec{S}//\vec{c}$ for Ver. A and Ver. B, respectively. All cuts along ($4/3,0,l$) have been broadened by a Lorentzian with a width that corresponds to the correlation length of a single trilayer.